\documentclass{svproc}
\usepackage{url}

\usepackage{graphicx}
\usepackage{amsmath}
\usepackage{amssymb}
\begin{document}
\mainmatter              
\title{Characterizing Distances of Networks on the Tensor Manifold}

\author{Bipul Islam\inst{1}, Ji Liu \inst{1}, Romeil Sandhu\inst{1}}

\institute{Stony Brook University, Stony Brook NY 08544, USA,\\
\email{romeil.sandhu@stonybrook.edu}}

\maketitle              

\begin{abstract}
At the core of understanding dynamical systems is the ability to maintain and control the systems behavior that includes notions of robustness, heterogeneity, and/or regime-shift detection.  Recently, to explore such functional properties, a convenient representation has been to model such dynamical systems as a weighted graph consisting of a finite, but very large number of interacting agents. This said, there exists very limited relevant statistical theory that is able cope with real-life data, i.e., how does perform analysis and/or statistics over a ``family'' of networks as opposed to a specific network or network-to-network variation.  Here, we are interested in the analysis of network families whereby each network represents a ``point'' on an underlying statistical manifold.  To do so, we explore the Riemannian structure of the tensor manifold developed by Pennec previously applied to Diffusion Tensor Imaging (DTI) towards the problem of network analysis. In particular, while this note focuses on Pennec definition of ``geodesics'' amongst a family of networks, we show how it lays the foundation for future work for developing measures of network robustness for regime-shift detection.  We conclude with experiments highlighting the proposed distance on synthetic networks and an application towards biological (stem-cell) systems.
\end{abstract}
\section{Introduction}
\begin{figure}[!t]
\begin{center}
\includegraphics[width=.9\textwidth]{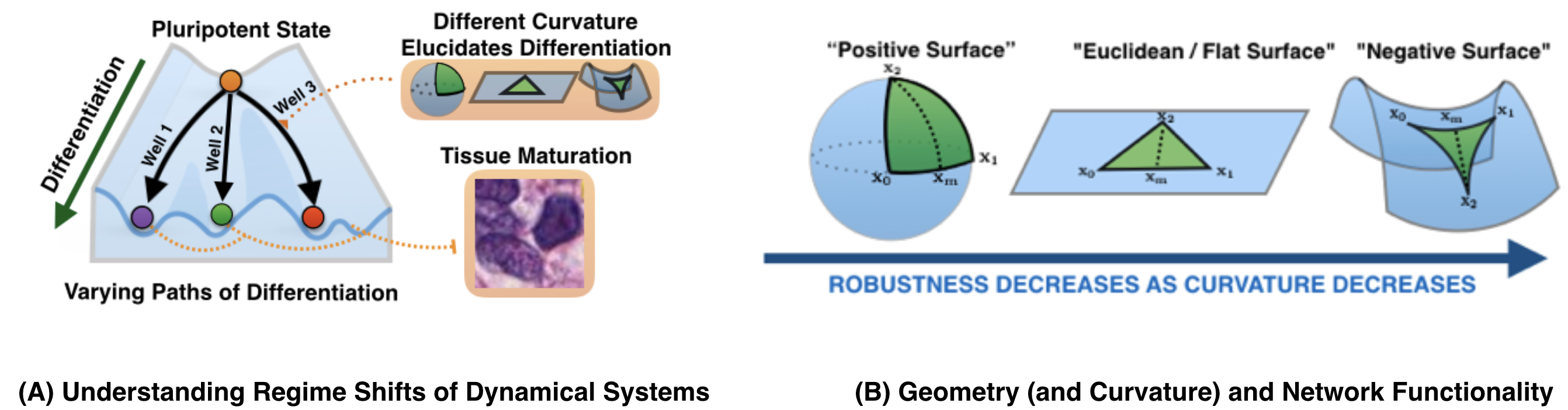}
\caption{This work focuses on the exploring a Riemannian framework  to elucidate regime-shifts and bifurcation detection in dynamical systems which can include, but not limited to cellular differentiation and signaling promiscuity in biological systems.}
\label{fig:sweet_spot}
\end{center}
\end{figure}

\begin{figure*}[!t]
\begin{center}
\includegraphics[width=1\textwidth]{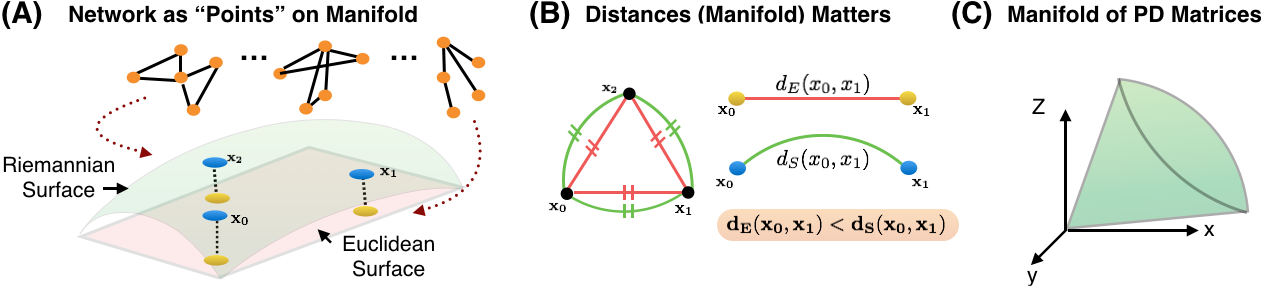}
\caption{(A) To understand global functionality, we treat networks as ``points'' on a manifold. (B) We require a framework for not only geodesic distance, but generalize distributions for network functionality \cite{Sandhu1}.  (C) Manifold of Positive Definite Matrices}
\label{fig:manifold}
\end{center}
\end{figure*}

Notions of robustness, heterogeneity, and phase changes are ubiquitous concepts employed in understanding complex dynamical systems with a variety of applications \cite{Bara1,Dem1,Tesch1,Bara2}. This is seen in Figure \ref{fig:sweet_spot}A. While recent advancements in network analysis has arisen through the usage of spectral techniques \cite{Chung}, expander graphs \cite{Widgerson}, percolation theory \cite{Aharony}, information theoretic approaches \cite{Dem1,Tesch1}, scale-free networks \cite{Bara2}, and a myriad of graph measures reliant on the underlying discrete graph space (e.g., degree distribution, shortest path, centrality \cite{Borgatti1,Bara3}), such measures rarely incorporate the underlying dynamics in its construction. To this end, we have previously developed fundamental relationships between network functionality \cite{Dem1,Varadhan} and certain topological and geometric properties of the corresponding graph \cite{Ollivier1,Ollivier2}. In particular, recent work of ours has shown the geometric notion of curvature (a measure of ``flatness'') is positively correlated with a system's robustness or its ability to adapt to dynamic changes \cite{Sandhu1,Sandhu2}.  This can be seen in Figure \ref{fig:sweet_spot}B.
Here, we are interested in expounding upon previous work \cite{Sandhu1} in order to exploit distances (and eventually statistics, robustness) over a family of networks. In particular, as vector-valued networks are becoming increasingly important, we believe such a framework may be of interest to broader complex network community.  To do so, we are interested in studying geometry of networks as a dynamical system that evolves over time in which each given network or observation may represent a ``point'' (encoded by \emph{some} matrix-based positive definite model) on an underlying Riemannian manifold.  From this, we are interested in applying the framework of Pennec \cite{Pennec1,Pennec2} to define geodesics such that distances between ``points'' (networks) in a given family can be measured as seen in Figure \ref{fig:manifold}.  This said, we note significant related work on employing statistical analysis for Riemannian manifolds has focused on directional statistics, statistics on shape spaces, as well as specific manifolds - we refer the reader \cite{Mardia,Moahker2,Smith} and references therein.   While these methods are mostly extrinsic whereby one embeds the manifold in an ambient Euclidean space, we are interested in studying \emph{intrinsic} Riemannian metrics that can be naturally applied to the network setting without restrictions on the network topology. As such, the remainder of the present paper is outlined as follows: In the next section, we provide preliminaries regarding the framework by Pennec \cite{Pennec1,Pennec2} primarily used for DTI imaging.  From this, we then show how this framework is a natural towards networks with a particular remarks on how we can begin on constructing robustness over network families.  Section 4 presents preliminary results on synthetic and biological data and Section 5 concludes with future work.
\section{Preliminaries}
This section focuses on developing a measures of distance amongst networks in the context of developing measures of network robustness (future work).
\subsection{Geodesics on the Network Manifold}
Let us consider a dynamical system that evolves over time and whose information is \emph{encapsulated by a positive definite matrix (tensor)}. In this regard, it is well-known that the space of tensors is not a vector space, but instead forms a Riemannian manifold $M$ \cite{DoCarmo}.  In particular, we will leverage the theory of symmetric spaces which has been extensively studied since the seminal work of Nomizu \cite{Nom}; a comprehensive work on tensor manifolds can be found in \cite{Joshi,Pennec1}. Riemannian manifolds are endowed with a metric that smoothly assign to each point $\rho\in M$ an inner product on $\mathcal{T_\rho}$, the tangent space to $M$ at $\rho$.  More formally, let us denote $\Upsilon$ and $\Upsilon^+$ and  as the set of all Hermitian matrices and the cone of positive-definite matrices, respectively. We consider the following:
\begin{align}
\begin{split}
&\Lambda_+ := \big\{ \rho \in \Upsilon^+ \quad \!\!\!\!\! | \quad \text{tr}(\rho)=1 \big \} \quad\quad \mathcal{T_\rho} := \big\{ \chi\in\Upsilon  \quad | \quad \text{tr}(\chi)=0 \big \}
\end{split}
\end{align}
where $\Lambda_+$ represents our space of networks and $\mathcal{T_\rho}$ is the corresponding tangent space.  From this, we let the observations of our dynamic system be recorded as $ \{\rho_0,\rho_1,\rho_2,...,\rho_t\}$ and where each ``point'' (network) $\rho_i\in{\Upsilon^+}$ is given by a tensor.  As seen in Figure \ref{fig:manifold}, we need a sensible notion of distance that is dependent on the underlying geometry. To do so, one defines the \emph{exponential map} as the function that maps to each vector $\overrightarrow{\rho_0 \rho_1}\in\mathcal{T}_{\rho_0}$, the point $\rho_1\in\Lambda^+$ on the manifold $M$ that is reached after unit time by the geodesic starting at $\rho_0$ with this tangent vector.  The exponential map, herein denoted as $\exp_{\rho}:\mathcal{T_\rho}\mapsto M$ at point $\rho$ is defined on the whole tangent space and for the particular space of tensors, this map is also one-to-one.  Moreover, one can also define a unique inverse map denoted as the \emph{logarithm map} $\text{Log}_{\rho_0} : M \mapsto \mathcal{T}_{\rho_0}$ that maps a point $\rho_1 \in M$ to the unique tangent vector $\chi\in\mathcal{T}_{\rho_0}$ at $\rho_0$ whose initial velocity is that of the unique geodesic $\gamma$  with $\gamma(0) = \rho_0$ and $\gamma(1)= \rho_1$.  Thus, the problem of computing distances amongst networks (encapsulated by a positive defense matrix) amounts to being able to properly computing geodesics on this space and for which, we must now consider a family of curves $\gamma(t)$ on the manifold and its speed vector $\dot\gamma(t)$.  Then, we seek the minimum distance (length) for any two points, e.g., $\rho_0$ and $\rho_1$ as
\begin{align}
\begin{split}
\mathcal{L}_{\rho_0}^{\rho_1} &= \min_\gamma \int_{\rho_0}^{\rho_1} \lVert \dot\gamma(t)\rVert_{\gamma(t)} dt = \min_\gamma \int_{\rho_0}^{\rho_1} \bigg(\langle \dot\gamma(t),\dot\gamma(t)\rangle_{\gamma(t)}\bigg)
\end{split}
\end{align}
where $\gamma(0)=\rho_0$ and $\gamma(1)=\rho_1$.  For this note, we relax the restriction upon trace one matrices for sake of clarity (without obfuscating the conceptual motivation).

\subsection{Revisiting Riemannian Framework for Tensors}
As highlighted, we seek to leverage Pennec \cite{Pennec1} framework towards the network setting.  While this framework has popularized most recently in the field of DTI \cite{Pennec2,Yogesh}, the mechanics can be used to analyze network functionality with this note focusing on the computation of distances amongst such networks.  In particular, the authors utilize a result from differential geometry regarding geodesics for invariant metrics on affine symmetric spaces \cite{Affine1} together with the Sylvester equation from control to propose an affine invariant metric whereby the unique geodesic curve from point $\rho$ (and at the origin) on our tensor (network) manifold with a tangent vector $\chi$ can be shown as
\begin{align}
\begin{split}
\gamma(t)_{(\rho,\chi)}&= \rho^{\frac{1}{2}}\exp(t\rho^{-\frac{1}{2}}\chi\rho^{-\frac{1}{2}})\rho^{\frac{1}{2}}\quad\text{and}\quad\gamma(t)_{(I,\chi)}= \exp(t\chi) 
\end{split}
\end{align}
where $\exp(\chi)=\sum_{k=0}^{+\infty}=\frac{\chi^k}{k!}$ is the usual matrix exponential.  From this, we have:
\begin{align}
\frac{d\gamma(t)}{dt}&= \frac{d}{dt}\exp(t\chi) \nonumber \\ 
&=\frac{d}{dt} \Bigg(Q \left[ {\begin{array}{cccc}
   e^{t\chi_{1,1}} & 0 & ... & 0 \\
   0 & e^{t\chi_{2,2}} & ... & 0 \\
    0 & 0 & ... & e^{t\chi_{n,n}}
  \end{array} } \right] Q^T\Bigg) \nonumber \\
  &= Q \text{DIAG}\big(\chi_{i}\exp(t\chi_{i})\big)Q^T \nonumber \\ 
  &= \exp(t\chi)^{\frac{1}{2}}\chi\exp(t\chi)^{\frac{1}{2}}) \nonumber \\ 
  &=\gamma(t)_{(I,\chi)}^{\frac{1}{2}}\star \chi
\end{align}
where the $A\star B=ABA^T$.  Noting the geodesic $\gamma(t)_{(\rho,\chi)}$ between any two points $\rho_0, \rho_1\in M$ where $\rho_1=\exp_{\rho_0}(\chi)$ at $t=1$, the \emph{logarithm map} can be seen as
\begin{align}
\gamma(1)_{(\rho_0,\chi)}&=\rho_1 = \rho_0^{\frac{1}{2}}\exp(\rho_0^{-\frac{1}{2}}\chi\rho_0^{-\frac{1}{2}})\rho_0^{\frac{1}{2}}\nonumber \\
\log(\rho_0^{-\frac{1}{2}}\rho_1\rho_0^{-\frac{1}{2}})&=\rho_0^{-\frac{1}{2}}\chi\rho_0^{-\frac{1}{2}} \nonumber\\
\rho_0^{\frac{1}{2}} \log(\rho_0^{-\frac{1}{2}}\rho_1\rho_0^{-\frac{1}{2}})\rho_0^{\frac{1}{2}} &=\chi = \overrightarrow{\rho_0\rho_1}=\text{Log}_{\rho_0}(\rho_1)
\end{align}
From this, the distance between tensors can finally be computed as
\begin{align}
\begin{split}
\text{dist}^{2}(\rho_0,\rho_1) &= \lVert \text{Log}_{\rho_0}(\rho_1) \rVert^{2}_{\rho_0} =\lVert \log(\rho_0^{-\frac{1}{2}}\rho_1\rho_0^{-\frac{1}{2}})\rVert^{2}_{2}.
\end{split}
\label{eq:riem_dist}
\end{align}
We now have a natural distance that measures the length of the geodesic curve that connects any two ``points'' on our tensor (network) manifold.  This is particularly important as it allows for characterization of functionality over a family of networks as we will highlight in the next sections.  
\subsection{Mean, Variance, and Gaussian Distributions}
Given the above geodesic distance, one can formulate necessary statistics and most importantly form distributions.  Specifically, with the above affine invariant metric, the unique mean $\mathbf{\bar{\rho}}$ of $N$ ``points'' $\{\rho_0,\rho_1,\rho_2,...,\rho_N\}$ on the tensor manifold can be computed via gradient descent:
\begin{equation}
\bar{\rho}_{t+1}=\bar{\rho}_t^{\frac{1}{2}}\exp\bigg(\frac{1}{N}\sum_{i=1}^{N}\log \Big(\bar{\rho}_t^{-\frac{1}{2}}\rho_{i}\bar{\rho}_t^{-\frac{1}{2}} \Big)\bigg)\bar{\rho}_t^{\frac{1}{2}}
\end{equation}
It has been noted that although no closed form expression for the mean exists, the above converges towards a steady-state solution in a few iterations.   As in the traditional Euclidean distance and similar to the Karcher or Frechet mean, the above mean is constructed on the premise that we seek a ``point'' that minimizes the sum of squared distance.e., $\rho=\sum_{i=1}^{N}\text{dist}^2(\rho,\rho_i)$.  In a similar fashion, the variance can be computed as follows:
\begin{equation}
\Sigma = \frac{1}{N-1}\sum_{i=1}^{N}\text{Vec}_{\bar{\rho}}\big(\text{Log}_{\bar{\rho}}(\rho_i)\big)\text{Vec}_{\bar{\rho}}\big(\text{Log}_{\bar{\rho}}(\rho_i)\big)^{T}
\end{equation}
where $\text{Vec}_{\bar{\rho}}(\rho_i)=\text{Vec}_{I}\big(\log(\bar{\rho}\star\rho_i)\big)$ and where the operation $\text{Vec}_I(A)$ is a vectorized projection of the independent coefficients of our tensor in the above formulation, i.e., $\text{Vec}_I(A) = (a_{1,1},\sqrt{2}a_{1,2},...\sqrt{2}a_{1,n},...\sqrt{2}a_{n-1,n}, a_{n,n})$. From this, one obtains a generalized form of the Gaussian distribution on the tensor manifold:
\begin{equation}
N_{(\bar{\rho},\Sigma)}(\rho_i)=k\exp\bigg(\frac{1}{2} \text{Vec}_{\bar{\rho}}\big(\text{Log}_{\bar{\rho}}(\rho_i)\big)^{T}\text{ }\Sigma^{-1} \text{ }\text{Vec}_{\bar{\rho}}\big(\text{Log}_{\bar{\rho}}(\rho_i)\bigg)
\end{equation}
With the above, we can now begin to stylize how such a framework can be utilized towards characterizing network families.
\section{Application To Network Characterization}
This section presents how the above framework can be used to generalize notions of robustness to a family of networks.  While the experiments focus on the fitness of ``distance'', we provide further information to put this note in context.
\subsection{Network Robustness From Scalar to Matrix Setting}
As highlighted, our main underlying interest is to understand the functionality over network families.  Recently, it has been
shown that Ricci curvature (herein denoted ``Ric'') from geometry is intrinsically connected to Boltzmann entropy $H$ \cite{LV,Sturm} as well as functional robustness of networks $R$ or the ability to maintain functionality in the presence of random fluctuations, i.e., $\Delta Ric\times \Delta R\geq 0$ \cite{Sandhu1,Sandhu2}.  Given that geometry (curvature) of networks seemingly plays an integral role in understanding functionality, we require an appropriate discretization applicable to a particular network.  To this end, one can discretely defined Ricci Curvature (due to Ollivier \cite{Ollivier1}) between any two ``points'' $x$ and $y$ as:
\begin{equation}
\kappa(x,y) := 1 - \frac{W_1(\mu_x,\mu_y)}{d(x,y)}
\label{eq:Ollivier}
\end{equation}
This definition, motivated by coarse geometry, is applicable to the graph setting whereby the geodesic distance $d(x,y)$ is given by the hop metric and $W_1$ can be computed simply via linear programming. However, this measure is limited to local network-to-network variation as opposed to a family of networks.  That is, the definition of equation (\ref{eq:Ollivier}) in the current setting has been used to analyze a given networks robustness through changes of curvature between any two end networks configurations.  Therefore, our primary motivation is focused on developing a matrix-valued curvature measure such that we can analyze Ricci curvature over a family of networks in a more global setting.  In particular, Ollivier's definition is rooted in the concept of providing an ``indicator'' between negative, Euclidean ``flat'', and positive curved spaces by measuring the ratio (distance) of the centers of a geodesic ball against that of the transport distance of ball's distributions (via Wasserstein-1).  Similarly, one can naturally expand this definition to the problem of a family of networks. To do so, we first need a geodesic distance $d_M(x,y)$ provided by equation (\ref{eq:riem_dist}) amongst networks along with the appropriate transport distance $W_p$.   For the case of Euclidean geometry, one can define the transport distance between two Gaussians ${\displaystyle \mu _{1}={\mathcal {N}}(m_{1},C_{1})}$ and ${\displaystyle \mu _{2}={\mathcal {N}}(m_{2},C_{2})}$, where $m_{1} $ and ${\displaystyle m_{2}\in \mathbb {R} ^{n}} {\displaystyle m_{2}\in \mathbb {R} ^{n}}$ are the means and $C_{{1}}$ and ${\displaystyle C_{2}\in \mathbb {R} ^{n\times n}} {\displaystyle C_{2}\in \mathbb {R} ^{n\times n}}$ are the covariances, as:
\begin{equation}
{\displaystyle W_{2}(\mu _{1},\mu _{2})^{2}=\|m_{1}-m_{2}\|_{2}^{2}+\mathop {\mathrm {trace} } {\bigl (}C_{1}+C_{2}-2{\bigl (}C_{2}^{1/2}C_{1}C_{2}^{1/2}{\bigr )}^{1/2}{\bigr )}.}
\end{equation} Comparing this definition to the structure presented in Section 2, the difference lies with respect to the underlying geometry for which Pennec \cite{Pennec1} provides the mathematical tools to generalize $W_2$ to our tensor manifold.  While the generalization is on-going and future work, this puts experimental results of $d_M(x,y)$ in context; e.g., our framework interest is to naturally generalize to functionality.
\subsection{Appropriate Matrix-Based Model: Graph Laplacian}
\label{sec:graph_lap}
Given our interest in network control, a natural network model for encapsulation is an approximated graph Laplacian whose structure provides a global representation of a given network with many useful properties that are intrinsically tied to system functionality.  While a complete review is beyond the scope of this note (we refer the reader to \cite {Chung}), a few areas that have garnered attention have included construction of expander graphs \cite{Widgerson} to Cheegers inequality with increasing attention on connections between spectral graph theory and its respective geometry \cite{Jost}.  Mathematically, one can define the Laplacian operator for a specific network as $\Delta f(x)= f(x) - \sum_y f(y)\mu_x(y)$ with $f$ being a real-valued function which coincides with the usual normalized graph Laplacian given by:
\begin{equation}
L = I - D^{-\frac{1}{2}}AD^{-\frac{1}{2}}
\end{equation}
It is interesting to note in this connection that if $k \leq \kappa(x,y)$ is a lower bound for Ricci curvature defined in equation (\ref{eq:Ollivier}), then the eigenvalues of Laplacian is bounded $k \leq \lambda_2 \leq . . . \leq \lambda_N \leq 2-k$ \cite{Jost}.  This relationship is important since $2-\lambda_N$ measures the deviation of the graph from being bipartite, i.e., a graph whose vertices can be divided into two disjoints sets $U$ and $V$ such that every edge connects a vertex in $U$ to one in $V$.  As such ideas appear in resource allocation and control, the study of such structures in the time-varying setting motivate our focus on the Laplacian matrix.  Lastly, while one can form a normalized Laplacian with unitary trace, we note that the Laplacian matrix by definition is a positive semi-definite matrix.  Therefore, to enforce positive definiteness, we utilized an approximated Laplacian $\hat{L}:= L+\epsilon_LI$ where $0<\epsilon_L << 1$ such that $x^{T}\hat{L}x>0$ for all nonzero $x\in \mathbb{R}^{n}$.  
\begin{figure*}[t]
\begin{center}
\includegraphics[width=1\textwidth]{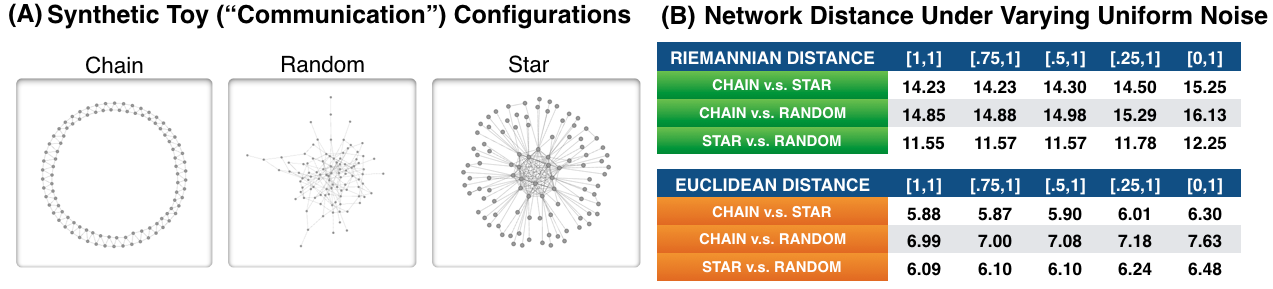}
\caption{(A) Graphical view of the three synthetic ``toy'' graphs that can be representative of varying types of communication networks, each with $200$ nodes and $400 edges$ (i.e., only topology changes).  (B) Distances of network-to-network under varying weights or ``noise.''   We see that the proposed framework is able to properly quantify difference as compared to Euclidean measure, i.e., the star-like network is ``closest'' to the random network as supported in \cite{Dem1,Sandhu1}.}
\label{fig:toy}
\end{center}
\end{figure*}
\section{Results}
We now present preliminary results to highlight potential for network analysis and set the foundation for future work in network robustness. 
\subsection{Toy Network Configurations}
To provide initial intuition of the proposed framework, we constructed three toy ``communication'' based networks.  As one can see in Figure \ref{fig:toy}, the networks are composed of $200$ nodes and $400$ edges with varying topology that represents a chain, a random, and a star-like network.  Similar to our earlier work \cite{Sandhu1}, we are interested in using the proposed framework to measure distances amongst these structures under varying ``noise''. As highlighted by network curvature \cite{Sandhu1} and network entropy \cite{Dem1}, one should expect from a communication perspective the star-like network is ``closer'' to a ``random'' network.  That is, given that their functional properties such as robustness (a measure of ``flatness'') are similar compared towards the chain-like network, we should expect that such networks on the manifold lie closer to one another.  This is exactly what is shown in Figure \ref{fig:toy}B.  Moreover, as we move from an unweighted graph to a weighted graph in which weights are chosen randomly in a uniform manner, this relationship still holds.  On the other hand, if we utilize the classical Euclidean distance (Frobenius norm) to measure network-to-network distance, we find that the chain-like network is in fact ``closer'' to the star-like network which is against intuition.
\subsection{Scale-Free and Random Networks}
\begin{figure*}[t]
\begin{center}
\includegraphics[width=1\textwidth]{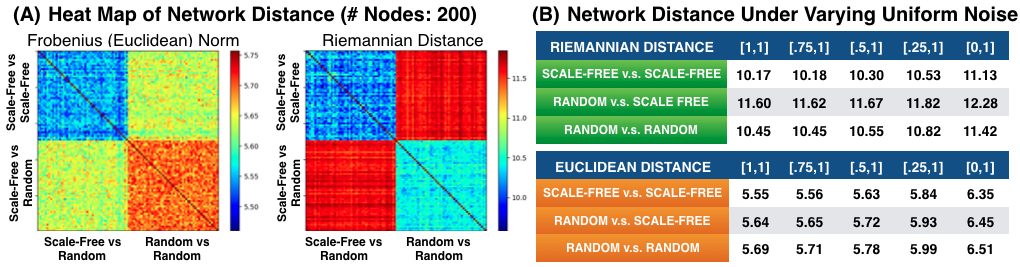}
\caption{Heat map of distances between scale-free and random networks measured by the proposed Riemannian metric to that of the Frobenius norm.  (B) Average distance of network-to-network under varying weights or ``noise.''   We see that the proposed framework is able to properly quantify difference as compared to Euclidean measure. Note:  Each network possess same number of nodes and edges, only topology changed.}
\label{fig:scale}
\end{center}
\end{figure*}
This section presents clustering results and importance of accounting for the underlying geometry.  In particular, we generate 50 scale-free networks \cite{Bara1} and 50 random (Erdos-Renyi) networks with each network composed 200 nodes and 1164 edges (differences are only in topology).  We then compute pairwise distances amongst all potential network pairs based on the Euclidean distance, defined by the Frobenius norm, as well as the Riemannian distance, defined by (\ref{eq:riem_dist}).  Figure \ref{fig:scale}A presents a ``heat map'' of the distances while Figure \ref{fig:scale}B presents average distance .  As one can see, Euclidean distance fails to cluster networks as the distance between scale-free to random networks are ``closer'' than random to random networks.   On the other hand, the framework proposed here is able to properly cluster the aforementioned networks which is an intuitive result given a basic understanding of geometry.  However, this motivates this framework for further study on network robustness.  We note if experiments are repeated and/or the number of nodes/edges are changed, the clustering result still holds.
\subsection{Waddingtons Landscape: Cell Differentiation}
\begin{figure*}[t]
\begin{center}
\includegraphics[width=1\textwidth]{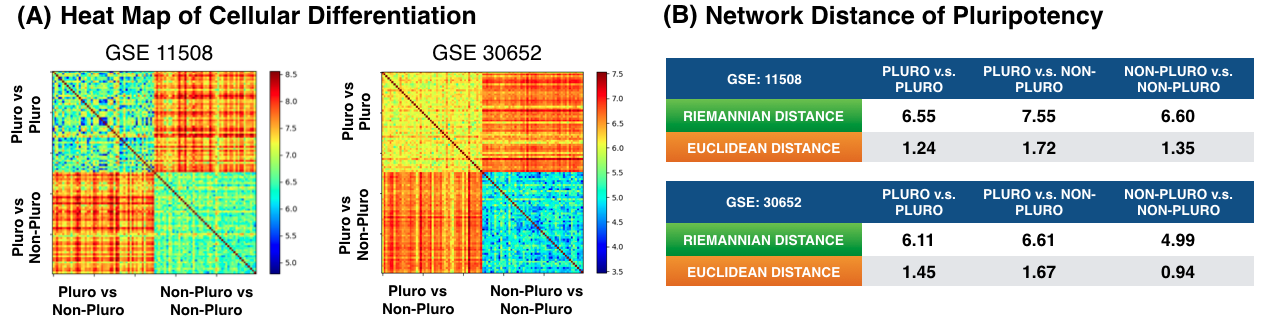}
\caption{Heat map of Riemannian distances between pluripotent and non-pluripotent networks under two public datasets.  (B) Average numerical distance between the pluro and non-pluro networks.  We see the proposed measure is able to classify such samples as well as the Euclidean measure.  We note the topology here is fixed across each networks.}
\label{fig:stem}
\end{center}
\end{figure*}

In this section, we apply the proposed distance measure towards cellular differentiation as depicted in Figure \ref{fig:sweet_spot}.  This epigenetic landscape provides a stylistic hierarchal view cellular differentiation, i.e., elevation associates to degrees of pluripotency or the ability for a cell to differentiate.  At a high level, one can consider the intricate and complex processes of cellular differentiation as a network whereby such bifurcations relate to differentiated states \cite{Tesch1}. To illustrate this, we accessed the Gene Expression Omnibus (GEO) of two public datasets, GSE11508 and GSE 30652, that possess 59 pluripotent / 160 non-pluripotent and 159 pluripotent / 32 non-pluripotent cell samples, respectively.  Within each pluripotent samples, there contains tissue types that include human embryonic stem cells (hESCs) and induced pluripotent stem cells (iPSCs).  This results in a potential mix of differentiation, i.e., iPSCs pluripotency may be much lower than hESCs yet are deemed pluripotent.  Here, we are interested to examine how ``far'' apart differentiated samples are from their counterparts.  As such, we construct a single pluripotent network similar to previous work \cite{Sandhu1} where the weights are the gene-to-gene correlation.  This was done similarly for a single non-pluripotent network.  Together, this was repeated to generate 50 pluripotent and non-pluripotent networks from each GSE11508 and GSE 30652.  Figure \ref{fig:stem} presents heat map results as well as the average network distance between such networks.  As one can see, we see a marked increase in distance between pluro and non-pluro samples.  We note while the Frobenius norm follows the same pattern, this is primarily due to the topology being fixed amongst samples as well as sparse interactions compared to work by Teschnedorff \cite{Tesch1}. We provide this to highlight the tacit topological dependence on distance which requires further study that is on-going (e.g., accounting for signaling interactome hierarchy).
 
\section{Conclusion}
This note introduces a statistical geometric framework \cite{Pennec1} previously applied to DTI Imaging \cite{Yogesh} toward the problem of network analysis.  To illustrate the importance of geometry in this setting, we presented preliminary results that show classical Euclidean distances are unable to properly differentiate between not only toy network configurations, but that of scale-free and random networks.  The results set the foundation for a more expansive study on global notions of robustness for which we will explore notions of network family curvatures.

 \end{document}